\documentclass[lettersize,journal]{IEEEtran}
\usepackage{amsmath,amsfonts}
\usepackage{algorithmic}
\usepackage{algorithm}
\usepackage{array}
\usepackage[caption=false,font=normalsize,labelfont=sf,textfont=sf]{subfig}
\usepackage{textcomp}
\usepackage{stfloats}
\usepackage{url}
\usepackage{verbatim}
\usepackage{graphicx}
\usepackage{cite}
\usepackage{fancyhdr} 
\begin{document}

\title{Cost-effective Network Disintegration through Targeted Enumeration}


\author{Zhigang Wang, Ye Deng, Petter Holme, Zengru Di, Linyuan L\"u, Jun Wu
\thanks{Zhigang Wang and Zengru Di are with the International Academic Center of Complex Systems, Beijing Normal University, Zhuhai 519087, China, and also with the School of Systems Science, Beijing Normal University, Beijing 100875, China.}
\thanks{Ye Deng and Jun Wu are with the International Academic Center of Complex Systems, Beijing Normal University, Zhuhai 519087, China (e-mail: junwu@bnu.edu.cn).}
\thanks{Petter Holme is with the Department of Computer Science, Aalto University, Espoo 02150, Finland, and also with the Center for Computational Social Science, Kobe University, Kobe 657-8501, Japan.}
\thanks{Linyuan L\"u is with the Institute of Fundamental and Frontier Sciences, University of Electronic Science and Technology of China, Chengdu 610054, China, and also with the Yangtze Delta Region Institute (Huzhou), University of Electronic Science and Technology of China, Huzhou 313001, China.}}

\markboth{\lowercase{\MakeUppercase{T}his work has been submitted to the \MakeUppercase{IEEE} for possible publication. \MakeUppercase{C}opyright may be transferred without notice, after which this version may no longer be accessible.}}%
{Shell \lowercase{\textit{et al.}}: A Sample Article Using IEEEtran.cls for IEEE Journals}

\IEEEpubid{}

\maketitle

\begin{abstract}
Finding an optimal subset of nodes or links to disintegrate harmful networks is a fundamental problem in network science, with potential applications to anti-terrorism, epidemic control, and many other fields of study. The challenge of the network disintegration problem is to balance the effectiveness and efficiency of strategies. In this paper, we propose a cost-effective targeted enumeration method for network disintegration. The proposed approach includes two stages: searching for candidate objects and identifying an optimal solution. In the first stage, we use rank aggregation to generate a comprehensive ranking of node importance, upon which we identify a small-scale candidate set of nodes to remove. In the second stage, we use an enumeration method to find an optimal combination among the candidate nodes. Extensive experimental results on synthetic and real-world networks demonstrate that the proposed method achieves a satisfying trade-off between effectiveness and efficiency. The introduced two-stage targeted enumeration framework can also be applied to other computationally intractable combinational optimization problems, from team assembly via portfolio investment to drug design. 
\end{abstract}


\section{Introduction}

\IEEEPARstart{E}{xploring} the internal correlation structure of complex networks is an important research paradigm for understanding complex systems\cite{wang2003complex, small2014random}. In most cases, we hope to ensure network connectivity, which has promoted research on network robustness in recent decades\cite{gao2011robustness, Schneider11, neff2021changes, Liu2020}. However, if a network is harmful, such as terrorist networks\cite{eiselt2018destabilization}, criminal networks\cite{calderoni2017communities}, epidemic spreading networks\cite{liu2021efficient}, financial contagion networks\cite{Kobayashi14} and cancer networks\cite{Quayle06}, efficiently disrupting the structure and function of the network becomes a meaningful and challenging task. This so-called network disintegration problem has attracted increasing attention among researchers\cite{2002Attack, wang2014damage, tan2016efficient, ren2019generalized, Kaiser2021}.

The core of the network disintegration problem---also known as the ``network attack,''~\cite{albert2000error, 2002Attack} ``graph fragmentation,''~\cite{aprile2019graph} and ``network dismantling''~\cite{braunstein2016network, ren2019generalized}---is determining the node or link set to be removed under certain constraints and various disintegration goals~\cite{ventresca2014derandomized, li2019bi}. This problem is typically NP-hard for general graphs~\cite{lalou2018critical} and its mathematical essence is a combinatorial optimization problem. In addition to early research based on exact combinatorial optimization methods to find an optimal network disintegration solution~\cite{wollmer1964removing, arulselvan2009detecting, veremyev2014exact}, researchers have also attempted to calculate the centrality measures of the nodes and then remove them individually, starting with the nodes with the highest centrality values, to develop a network disintegration strategy~\cite{albert2000error, 2002Attack, lu2016vital}. However, the set composed of a single important node may not be the most critical set of nodes, and with the increased availability of large-scale networks, novel heuristic or approximate algorithms have been proposed to find vital nodes in complex networks~\cite{morone2015influence, mugisha2016identifying, zdeborova2016fast}. A recent study suggested an iterative algorithm to select multiple controlled nodes based on the spectral properties of the grounded Laplacian matrix obtained by deleting specific rows and columns from the Laplacian matrix of the network~\cite{liu2021optimizing}. Furthermore, some studies introduced evolutionary algorithms to the network disintegration problem and attempted to find a near-optimal strategy from the considerable solution space~\cite{lozano2017optimizing, deng2018optimal}. Inspired by advances in artificial intelligence to solve many practical problems, some studies have developed deep reinforcement learning or machine learning to find influential nodes in complex networks~\cite{fan2020finding, 2021Machine}.

An outstanding challenge in the network disintegration problem is to take into account the computational cost. Although considerable progress has been made in the study of network disintegration, it remains challenging to achieve a good balance between effectiveness and efficiency. Methods with good effectiveness (giving a more accurate estimate), such as mathematical programming, evolutionary algorithms, and deep learning approaches, typically have poor efficiency (effectiveness per running time), limiting their applications in large-scale networks. On the contrary, high-efficiency methods, such as centrality methods and heuristic algorithms, are typically unsatisfactory in terms of effectiveness, yielding nonoptimal solutions. Similar tradeoff problems have been studied for related tasks to estimate the highest degree and the optimal individuals to vaccinate~\cite{holme2017cost}. To find a compromise between effectiveness and efficiency, we propose a targeted enumeration method in this paper. We first extract a small-scale candidate set of nodes to reduce the scope of the enumeration, and then find the optimal combination among the candidate nodes through enumeration. The core and difficulty of the method is to efficiently determine the set of candidates. We propose solving this problem using rank aggregation. The resulting two-stage targeted enumeration method has a highly flexible framework that does not require domain-specific knowledge and leads to a cost-effective network disintegration strategy.

\section{Network disintegration model}

Consider an undirected and unweighted graph $G=(V, E)$ with a finite set of nonempty nodes $V$ and a set of links $E$. Let $N=|V|$ and $W=|E|$ be the number of nodes and links, respectively, and define different nodes as $1,2,\cdots ,N$. This paper focuses on the removal of nodes and assumes that all links connected to the node will be deleted after the node is removed. Let $\hat V \subseteq V\ $ denote the set of nodes to be removed; thus, $\hat G = (V - \hat V,\hat E)$ is the network that remains after removing the nodes in $\hat V$, and $n = | {\hat V} |$ is the strength of disintegration. As a reference, let $\tilde{G}$ be the residual network after randomly removing the $n$ nodes. We denote the network disintegration strategy as $X = [{{x_1},{x_2},\cdots,{x_N}} ]$, and its elements are $x_{i} = 1$ if the corresponding $i$th node satisfies $i\in \hat{V}$; otherwise, $x_{i} = 0$; thus, we can obtain the disintegration strength as $n=\sum\nolimits_{i=1}^N{x_i}$. Regardless of the type and scale of attacks to which the network is subject, it will inevitably damage its inherent structure and functions, which will also be reflected in the objective function of the network performance. Based on this, we introduce the following objective function to measure the disintegration effect
\begin{equation}\label{objective_function}
	{\Phi (X)=\frac{\varGamma (G)-\varGamma (\hat{G})}{\varGamma (G)-\varGamma (\tilde{G})},} 
\end{equation}
where $\varGamma$ represents the measurement function of network performance. If $G_1=\left( V_1, E_1 \right)$ is a proper subgraph of $G_2=\left( V_2, E_2 \right)$, that is, $V_1\subset V_2$ or $E_1 \subset E_2$, we assume that $\varGamma \left( G_1 \right) < \varGamma \left( G_2 \right)$. The monotonicity of $\varGamma$ ensures that the network performance strictly decreases monotonically with the network disintegration process and leads to $\Phi>0$ if $n>0$. $\Phi$ reflects the disintegration effect of different network disintegration strategies. The larger $\Phi$ suggests a better disintegration effect. There is an important reference value, that is, $\Phi=1$. If $\Phi > 1$, it means that the disintegration strategy is superior to random removal of nodes. Eq.\eqref{objective_function} shows that the goal is to design a node removal strategy, that is, a subset of nodes to be removed, which can maximize the disintegration effect $\Phi$. Thus, the optimization model for the disintegration strategy can be described as the following general mathematical model
\begin{equation}\label{optimization_model}
	\begin{aligned}
		&\max \Phi (X = [ x_1,x_2,\cdots,x_N ]),\\
		&\mathrm{s.t.} ~
		\begin{cases}
			\sum_{i=1}^N{x_i}=n \\[12pt]
			x_i = 0 \ \mathrm{or}\ 1,&i = 1,2,\cdots,N.
		\end{cases}
	\end{aligned}
\end{equation}

Usually, the disintegration effect is measured by the size of the largest connected component~\cite{albert2000error}. However, it changes very little when removing a small number of nodes from the network. Therefore, in this study, we employ natural connectivity~\cite{jun2010natural, wu2011spectral} as a measure function $\varGamma$ among a variety of alternative ways. Natural connectivity is a measure function of structural robustness in complex networks, which can be mathematically derived from the graph spectrum~\cite{jun2010natural, wu2011spectral}. This measure characterizes the redundancy of alternative links by weighting the total number of closed walks with all lengths in the network and can also be interpreted as the Helmholtz free energy of a network~\cite{estrada2012physics}. From a mathematical perspective, it can be derived from the graph spectrum as an average eigenvalue:
\begin{equation}\label{natural_connectivity}
	{NC = \ln \!\left( {\frac{S}{N} }\right) = \ln \!\left( {\frac{1}{N}\sum_{i = 1}^N {{e^{{\lambda _i}}}} } \right),}  
\end{equation}
where $S$ is the total weighted number of closed walks and $A(G)$ is the adjacency matrix of the network $G$ with eigenvalues ${\lambda _1}\geq{\lambda _2}\geq \cdots \geq {\lambda _N}$, which is called the spectrum of $G$. 

Natural connectivity has been shown to change strictly monotonically with the addition or deletion of links and then provides a sensitive and reliable measure of the robustness of the graph~\cite{jun2010natural, wu2011spectral, wu2012robustness, wu2011robustness}. Moreover, for networks with a large spectral gap between the largest eigenvalue $\lambda_1$ and the second largest eigenvalue, we can consider the following approximation of natural connectivity~\cite{tan2016approximating}:
\begin{equation}
	{NC=\ln \!\left[ \left( \sum_{i=2}^N{e^{\lambda _i}+e^{\lambda _1}} \right) /N \right] \approx \lambda _1-\ln N.}
\end{equation}

\section{Searching the candidate objects by rank aggregation}

From a mathematical perspective, network disintegration is a typical combinatorial problem that considers $n$ nodes from $N$ nodes without repetition. For a small network size $N$, we can directly obtain an optimal solution by enumerating all combinations of $C_{N}^{n}=\frac{N!}{n!(N-n)!}$. However, for large-scale networks, there will be a problem of combinatorial explosion. To construct a heuristic method, the selected $n$ nodes should be important according to some criterion. If we extract a small-scale candidate set of vital nodes $\tilde{V}$ in advance and then enumerate all combinations only among the candidate set, it will dramatically improve the efficiency of the enumeration. We use $\tilde{N}$ to denote the size of the candidate set $\tilde{V}$, where $n \leq \tilde{N} \leq N $. Then, the enumeration range can be reduced from $C_{N}^{n}$ to $C_{\tilde{N}}^{n}$. Now, the core problem is to find candidate objects. There are numerous criteria that characterize the importance of nodes. If we only use a single criterion, then some potential key nodes may be missed. Therefore, we simultaneously consider multiple node importance criteria using rank aggregation (RA). In network science, the centrality of nodes is a common approach to assess the importance of nodes~\cite{lu2016vital}. Thus, we first generate multiple node rankings based on various centrality measures. Then we combine these individual rankings into a consensus ranking using the rank aggregation method. Finally, we determine the candidate objects $\tilde{V}$ based on the consensus ranking. 

In this study, among a variety of alternative methods, we choose the graph-based rank aggregation method~\cite{xiao2019graph, zhang2021comprehensive} to aggregate these individual rankings into a single consensus ranking $\hat{R}$. The graph-based rank aggregation method has been shown to outperform other rank aggregation methods, particularly for high-dimensional ranking. 

Consider $M$ rankings of $N$ nodes given by the $M$ node importance criterion and use $R_i=[r_{i1}, r_{i2}, \cdots, r_{iN}]$ to denote the node importance ranking given by the criterion $c_i$, where $r_{ij}$ represents the rank of the node $j$ based on the criterion $c_i (i=1,2,\cdots,M)$. The transition matrix for the criterion $c_i$ is denoted by $P^{c_i}=(p_{st}^{c_i})_{N\times N}$, where $p_{st}^{c_i}=1$ if node $s$ outranks node $t$ under $c_i$; otherwise, $p_{st}^{c_i}=0$. Based on the transition matrix, we denote the adjacency matrix for a competition graph as $A=(a_{st})_{N\times N}$ , where $a_{st}=\sum\nolimits_{i=1}^M{p_{st}^{c_i}}$. Furthermore, based on the adjacency matrix $A$, we denote the competition graph of the network nodes as $G_c$. The nodes in the directed and weighted graph $G_c$ represent the nodes in the real network, and each directed link $e_{st}$ represents an outranking relation from node $s$ to $t$. The weight of the directed link $e_{st}$ represents the number of times node $s$ is placed ahead of node $t$ in all aggregated measure rankings. We also denote the in-degree and out-degree of node $j$ in the competition graph $G_c$ by $d_{j}^{-}=\sum\nolimits_{s=1}^N{a_{sj}}$ and $d_{j}^{+}=\sum\nolimits_{t=1}^N{a_{jt}}$, respectively. Thus, we can define the ratio of out-in degrees (ROID) as follows:
\begin{equation}
{\alpha _j=\frac{d_{j}^{+}+1}{d_{j}^{-}+1}}, 
\end{equation}
which can be used to quantify the strength of node $j$ and rank all nodes according to their ROID~\cite{xiao2019graph}. The higher the ROID value, the higher the rank of the nodes.

\begin{figure*}
\centering
\includegraphics[width=1\textwidth]{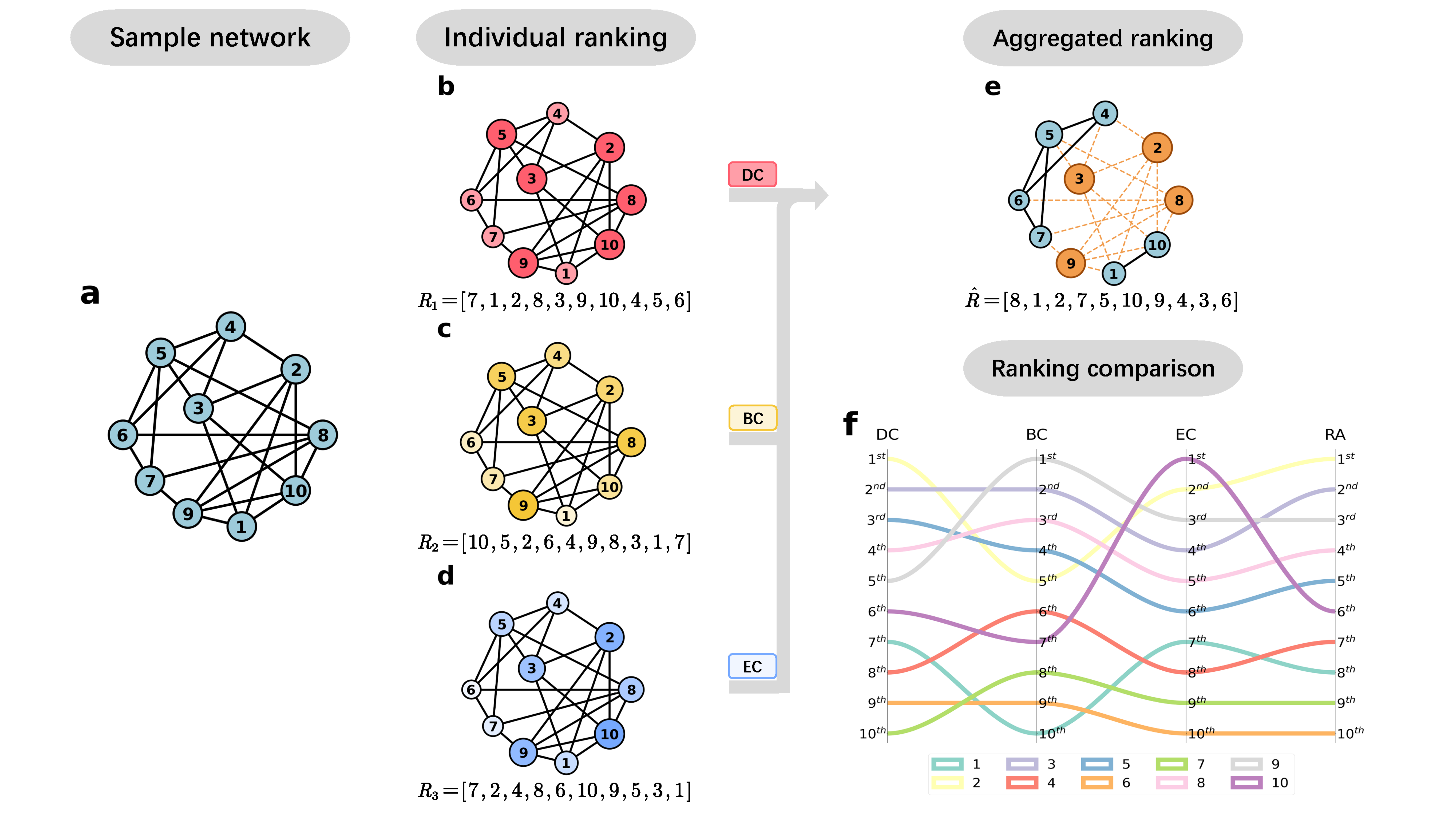}
\caption{Illustration of searching the set of candidates by aggregating the rankings. (a) The sample network, where $N=10$, $W=23$, $n=2$, and $\tilde{N}=4$. And the numbers represent the labels of the nodes. (b) to (d) Individual node rankings based on degree centrality, betweenness centrality, and eigenvector centrality, respectively. The size of the node is proportional to its ranking. (e) The aggregated ranking of the nodes. The orange nodes make up a set of candidates $\tilde{V}$. (f) Comparison of the ranking of nodes with various centrality measures.}
\label{fig:1}
\end{figure*}

To better understand the process of searching for candidate objects, an illustration is shown in Fig.~\ref{fig:1}. Taking into account a sample network that contains 10 nodes and 23 links and has a network topology as shown in Fig.~\ref{fig:1}(a), we employ three common centrality measures: degree centrality (DC)~\cite{albert2000error}, betweenness centrality (BC)~\cite{2002Attack}, eigenvector centrality (EC)~\cite{roddenberry2021blind}. The individual ranking of the nodes based on the three centrality measures is shown in Fig.~\ref{fig:1}(b), (c), and (d). The aggregated ranking $\hat{R}$ is shown in Fig.~\ref{fig:1}(e). Details on the ranking are provided in Table ~\ref{tab:1}. We set the disintegration strength $n$ as 2 and the size of the candidate set $\tilde{N}$ as 4 and then obtain the candidate set $\{2, 3, 8, 9\}$ based on the aggregated ranking, as shown in the orange node in Fig.~\ref{fig:1}(e). The comparison results of the node ranking with different centrality measures are visualized in Fig.~\ref{fig:1}(f). Each curve represents a node, and the height of the curve represents the node ranking according to the corresponding criterion. The wavy curves suggest that there are distinct differences between the three individual rankings. For example, node $2$ ranks first with DC but fifth with BC; node $10$ ranks first with EC but sixth with DC. In the far right of Fig.~\ref{fig:1}(f), the aggregated ranks are also presented. The RA method integrates all information from individual rankings and achieves a comprehensive ranking, effectively overcoming the one-sidedness of the individual measure. To some extent, this method takes the ``average'' of multiple rankings.

\begin{table}[htbp]
\centering
\setlength{\tabcolsep}{4pt}
\caption{The rankings and values of nodes in the sample network based on different centrality measures and rank aggregation}
\label{tab:1}
\begin{tabular}{ccccccccc}
      \hline
      Node ID   & \multicolumn{2}{c}{DC}    & \multicolumn{2}{c}{BC}
	  & \multicolumn{2}{c}{EC}     & \multicolumn{2}{c}{RA}    \\ \cline{2-9} 
	  & \multicolumn{1}{c}{Rank}   & \multicolumn{1}{c}{Value} &  \multicolumn{1}{c}{Rank} 
	  & \multicolumn{1}{c}{Value}  & \multicolumn{1}{c}{Rank}  & \multicolumn{1}{c}{Value} 
	  & \multicolumn{1}{c}{Rank}   & \multicolumn{1}{c}{ROID}   \\ \hline
		$1$     & 7   & 4   & 10  & 0.0093 & 7   & 0.3002  & 8   & 0.3182   \\ 
		$2$     & 1   & 5   & 5   & 0.0694 & 2   & 0.3483  & 1   & 3.8333   \\  
		$3$     & 2   & 5   & 2   & 0.0926 & 4   & 0.3431  & 2   & 3.8333   \\
		$4$     & 8   & 4   & 6   & 0.0648 & 8   & 0.2717  & 7   & 0.4500   \\
		$5$     & 3   & 5   & 4   & 0.0880 & 6   & 0.3165  & 5   & 1.6364   \\
		$6$     & 9   & 4   & 9   & 0.0278 & 10  & 0.2557  & 10  & 0.1154   \\
		$7$     & 10  & 4   & 8   & 0.0324 & 9   & 0.2689  & 9   & 0.1600   \\
		$8$     & 4   & 5   & 3   & 0.0926 & 5   & 0.3324  & 4   & 1.9000   \\
		$9$     & 5   & 5   & 1   & 0.1065 & 3   & 0.3459  & 3   & 3.1429   \\
		${10}$  & 6   & 5   & 7   & 0.0556 & 1   & 0.3591  & 6   & 1.4167   \\
		\hline                    
\end{tabular}
\end{table}

\begin{figure*}
\centering
\includegraphics[width=0.95\textwidth]{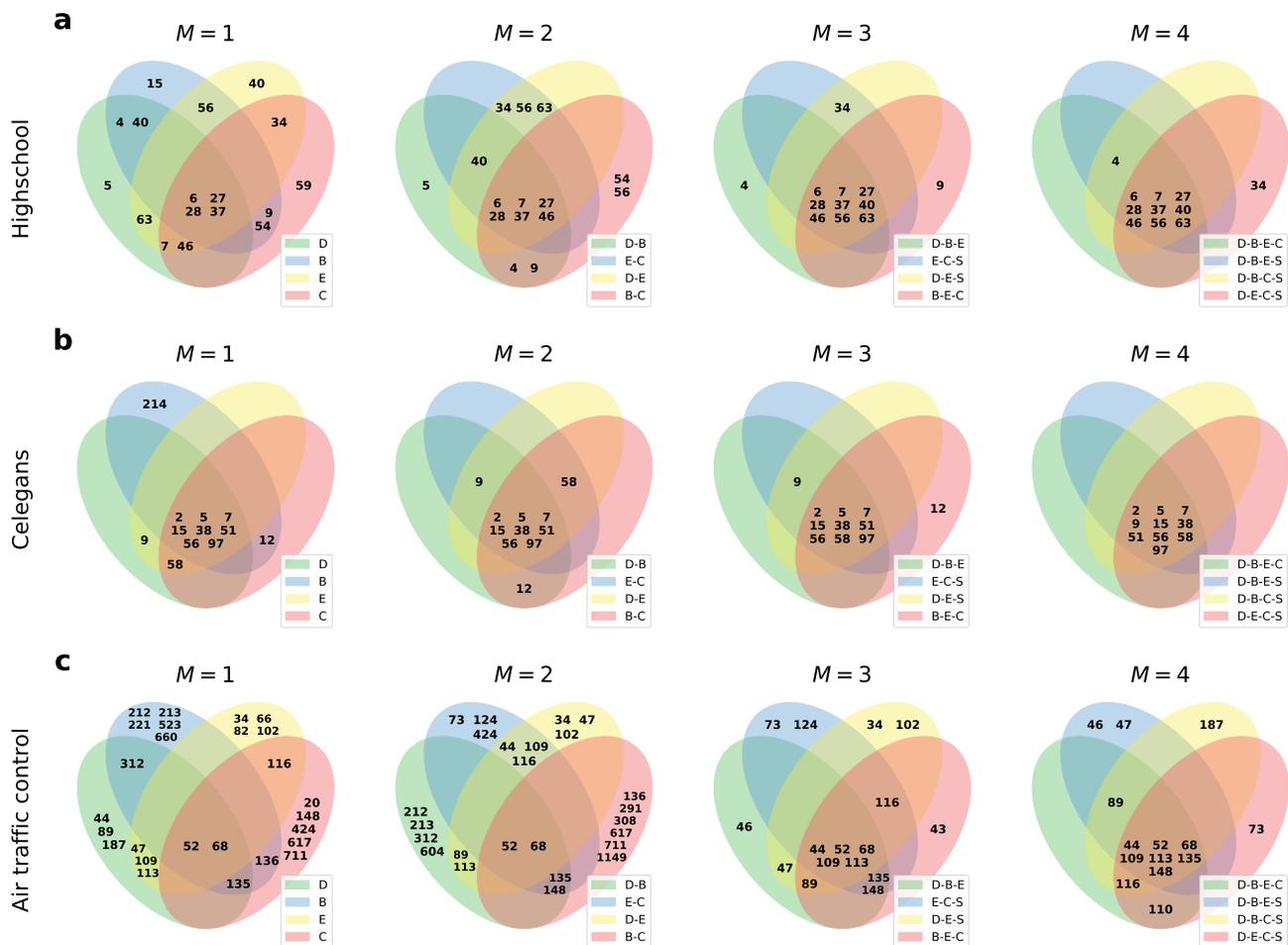}
\caption{The Venn diagram of candidate sets based on various combinations of node importance criteria in real-world networks. In the figure, D, B, E, C, and S represent the degree centrality, betweenness centrality, eigenvector centrality, closeness centrality, and subgraph centrality, respectively. The size of the candidate set $\tilde{N}$ is 10. (a) The network contains friendships between boys in a small high school in Illinois, where a node represents a boy and an link between two boys shows that they are friends. And the numbers represent the labels of the nodes. (b) The metabolic network of \textit{Caenorhabditis elegans}. In this representation, a metabolic network is made up of nodes, substrates that are connected to each other through links, which are the actual metabolic reactions. (c) In the air traffic control network, the nodes represent airports or service centers, and links are created from the preferred routes recommended by the National Flight Data Center.}
\label{fig:2}
\end{figure*}

Intuitively, the number of criteria for the importance of the node $M$ and the combination of these criteria will affect the candidate objects and further influence the disintegration effect. To explore the effect of the node importance criterion on the candidate set $\tilde{V}$, Fig.~\ref{fig:2} shows the Venn diagram of candidate sets obtained using various combinations of node importance criteria in three real-world networks. As we see in Fig.~\ref{fig:2}, if we only use a single criterion ($M=1$), the set of candidates with different combinations of criteria varies significantly. However, as $M$ increases, the intersection of candidate sets with different criteria also expands observably. For example, in the network shown in Fig.~\ref{fig:2}(a), there are only 4 overlapping nodes when $M=1$ but 9 overlapping nodes when $M=3$; these results indicate that rank aggregation can help us search for a stable and credible candidate set. Without loss of generality, we choose D-B-E as the combination of the node importance criterion in the following experimental analysis.

\section{Identifying the optimal solution by targeted enumeration}

In the previous section, we proposed selecting $\tilde{N}$ candidate nodes by rank aggregation. Now, we need to find the optimal combination among the candidate set through enumeration. The size of the candidate set $\tilde{N}$ will directly affect the effectiveness and efficiency of the proposed method. Considering that $n \leq \tilde{N} \leq N$, we assume that $\tilde{N}=n+(N-n)\alpha$, where $0 \leq \alpha \leq 1$ is the redundancy coefficient. When $\alpha$ reaches the maximum value 1, it becomes an exhaustive enumeration. While $\alpha < 1$, we call it targeted enumeration (TE).

\begin{figure*}
\centering
\includegraphics[width=.95\textwidth]{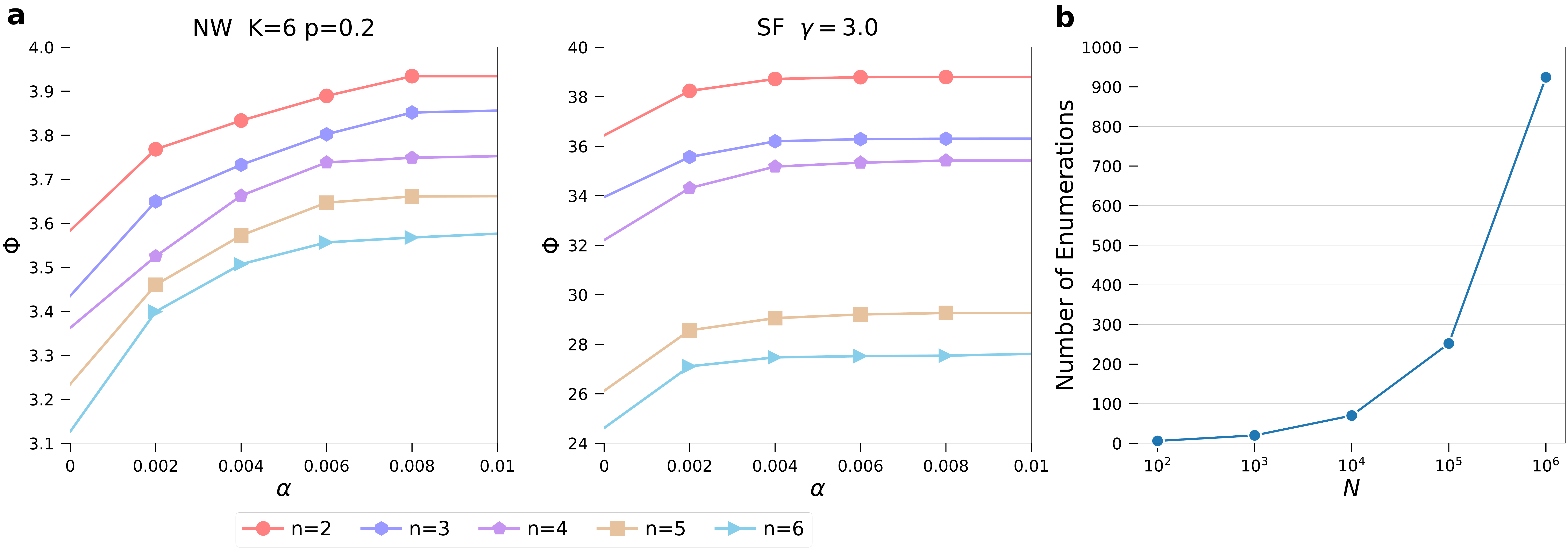}
\caption{The effectiveness and efficiency of the TE method. (a) The network disintegration effect $\Phi$ under different redundancy coefficients $\alpha$. The results shown are the average of 10 network instances under the same parameters. Results for the NW network of size $N=1000$. The NW network starts with a regular network with local connections in the range $K=6$, with the probability $p=0.2$ of adding a new link between a randomly selected unconnected pair of nodes. Scale-free network with size $N=1000$ and degree exponent $\gamma=3.0$. (b) The number of enumerations as a function of network size $N$ according to Eq.\eqref{enumeration}  when assuming $n = \log (N)$ and $\alpha = \log (N)/N$.}
\label{fig:3}
\end{figure*}

A higher $\alpha$ will lead to better effectiveness but worse efficiency. Fig.~\ref{fig:3}(a) shows the disintegration effect $\Phi$ as a function of the redundancy coefficient $\alpha$ in two typical synthetic networks: the Newman-Watts (NW) model of small-world network\cite{watts1998collective}, and the scale-free (SF) network\cite{barabasi1999emergence}. The curve shown first increases and then flattens, indicating that a small value of the redundancy coefficient is sufficient for the targeted enumeration and increasing $\alpha$ contributes little to the disintegration effect. These results also suggest that the process of selecting candidate objects is effective to some extent. In practical applications, the value of $\alpha$ can be determined based on real needs. 

The algorithmic process of the TE is summarized below. First, we choose $\tilde{N}$ candidate nodes based on the aggregate ranking of the nodes. Then, we enumerate all possible combinations among the candidate set. Finally, we find the optimal solution that corresponds to the largest disintegration effect $\Phi$. In the example shown in Fig.~\ref{fig:1}, if the redundancy coefficient is considered to be $\alpha=0.25$, then there are $C_{\tilde{N}}^{n}=C_{4}^{2}=6$ combinations, among which the combination $\{2,8\}$ is the optimal solution.

Next, we briefly analyze the time complexity of the TE method. As described above, the time complexity of the TE method includes three parts: calculating the centrality of the nodes, aggregating multiple rankings, and enumerating among the candidate sets. In the first part, the time complexity for DC is $O(W)$, the time complexity for BC is $O(NW)$~\cite{brandes2001faster}, and the time complexity for EC is $O(N+W)$~\cite{bonacich1972factoring}. In the second part, the time complexity of the rank aggregation is $O(\tilde{N}^2)$. Considering that $n \ll N$ and $\alpha \ll 1$ in most realistic cases, we can also assume that $n = \log (N)$ and $\alpha = \log (N)/N$ and then obtain $\tilde{N}=n+(N-n)\alpha \approx  2\log(N)$. Thus, the time complexity of the second part is $O(\log ^2(N))$. In the third part, with the assumption that $n = \log (N)$ and $\alpha = \log (N)/N$, the number of enumerations can be given as: 
\begin{equation}\label{enumeration}
   C_{\tilde{N}}^{n} = C_{2\log N}^{\log N}=\frac{\left(2\log N\right)!}{\left[\left(\log N\right)!\right]^2}.
\end{equation}
A schematic of the enumeration times $C_{\tilde{N}}^{n}$ with a varying network size $N$ when assuming $n = \log (N)$ and $\alpha = \log (N)/N$ is shown in Fig.~\ref{fig:3}(b). We see that the number of enumerations is less than 1000, even with the large network size $N=10^6$, which is acceptable. 

\section{Experimental analysis on synthetic and real-world networks}

\subsection{Experiments in synthetic networks}

To demonstrate the applicability of the proposed method, we next evaluate its performance on two kinds of typical synthetic networks: the NW network and the SF network. We use five other methods for comparison: degree centrality, betweenness centrality, eigenvector centrality, collective influence (CI)~\cite{morone2015influence} and tabu search (TS)~\cite{deng2019optimal}.  

\begin{figure*}
\centering
\includegraphics[width=.88\textwidth]{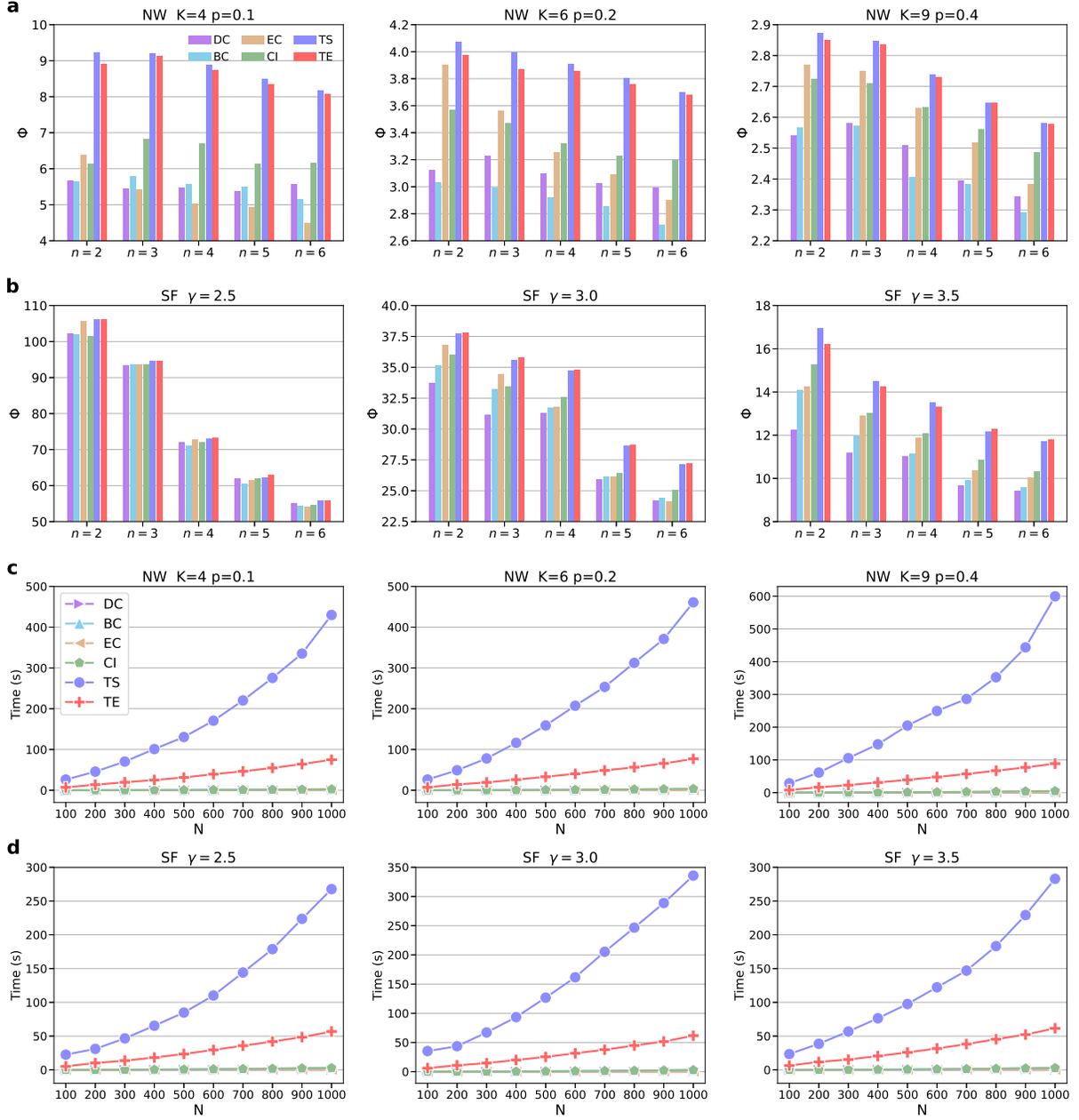}
\caption{Performance of TE in synthetic networks. We set the geodesic distance $\ell$ as 2 for the CI method. For the TS algorithm, we assign the tabu list length to 5, the number of candidate solutions to 5, and the maximum number of iterations without improving the optimal solution will be 2000. The numerical results shown are the averages of 20 different network instances under the same parameters. (a) The disintegration effect of the TE method on NW network with size $N=1000$ of varying neighbor numbers $K$ and connection probability $p$. (b) The disintegration effect of the TE method on SF network with size $N=1000$ of varying degree exponent $\gamma$. (c) The computation time of different methods as a function of network size $N$ on the NW network. All simulation results are obtained on a desktop computer with an Intel Core i7-9700 CPU with 3.00GHz and 16.0 GB of RAM. (d) The computation time of different methods on the SF network.}
\label{fig:4}
\end{figure*}

Fig.~\ref{fig:4}(a) and (b) show the disintegration effect $\Phi$ as a function of the disintegration strength $n$ with different disintegration methods. We also set $\alpha$ equal to 0.01. As shown in Fig.~\ref{fig:4}(a) and (b), the proposed method is almost close to the TS method, which can achieve a good disintegration effect. Both methods consistently outperform other methods on all synthetic networks. It is worth pointing out that, even for the heterogeneous SF network with $\gamma=2.5$, in which the vital nodes are apparent and then all methods work well, the TE method still maintains a weak advantage compared to other methods except for the TS method. In addition to improved effectiveness, the TE is also markedly efficient. Fig.~\ref{fig:4}(c) and (d) show the computation time of different methods as a function of network size. As shown in Fig.~\ref{fig:4}(c) and (d), with increasing network scale, the growth rate of the TS method is markedly higher than that of the other methods. In contrast, the proposed method is more efficient. 

\subsection{Experiments in real-world networks}

\begin{figure*}
\centering
\includegraphics[width=.85\textwidth]{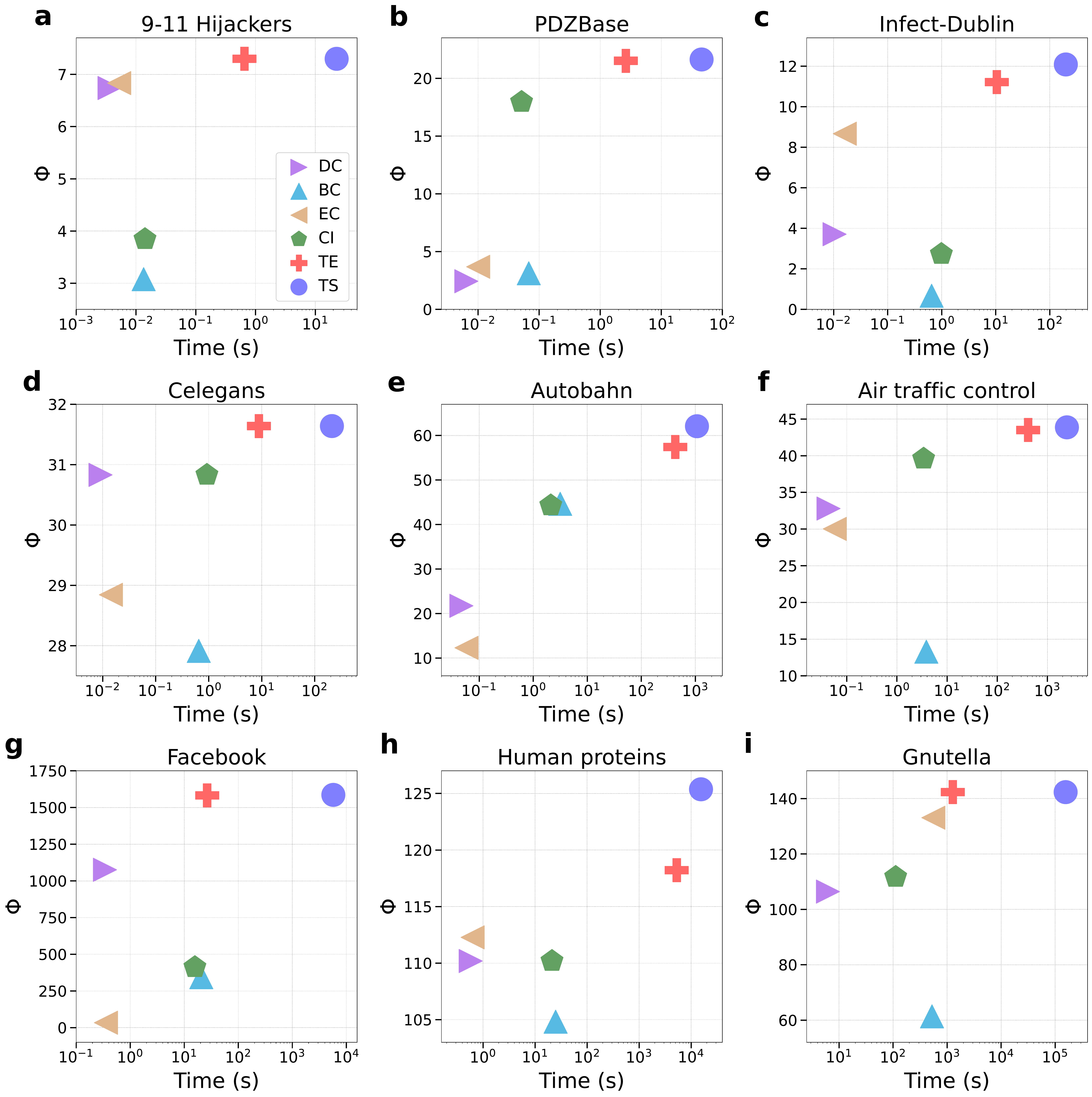}
\caption{Performance of TE in real-world networks. We evaluated the disintegration effect $\Phi$ and the running time of the six methods on nine real-world networks of different types and set the disintegration strength as $n=\ln N$ for different networks.}
\label{fig:5}
\end{figure*}

Since synthetic networks cannot completely summarize the typical properties of real-world networks, we apply the TE method to several realistic scenarios using the aforementioned methods. Table ~\ref{tab:2} shows details of real-world networks used in our study. The data sets are publicly accessible and are retrieved from the KONECT Project (http://konect.cc/), the Network Data Repository (https://networkrepository.com/index.php), and the Colorado Index of Complex Networks (https://icon.colorado.edu). We assume that the real-world networks considered in this paper are simple graphs with undirected, unweighted, and single links. We show the disintegration effect $\Phi$ and the running time of the six methods in Fig.~\ref{fig:5}. Along with the TS method, the proposed method achieves superior performance compared to the other four methods with respect to the disintegration effect. It is obvious that the disintegration effect of these two methods is more stable. For example, for the disintegration strategy based on EC, its effect is second only to TE and TS methods in 9-11 Hijackers, Infect-Dublin and Gnutella networks, but not so good in Autobahn and Facebook networks. However, the TS method leads to good effectiveness but poor efficiency. In other words, the proposed method has lower cost to obtain a disintegration effect that is similar to that achieved by the TS method. Compared to centrality-based methods, although the efficiency of the proposed method is lower than that of centrality-based methods, it is acceptable, indicating that the proposed method achieves a satisfying balance between effectiveness and efficiency. 

\begin{table*}[htbp]
\centering
\caption{Details of the real-world networks}
\label{tab:2}	
\resizebox{\linewidth}{!}{
\begin{tabular}{lllllll}
\hline
Name & Number of nodes & Number of links & Category & Network format & Node meaning & Edge meaning \\ 
\hline 
9-11 Hijackers & 62 & 304 & Terrorist network & Undirected & Person & Association \\
PDZBase & 212 & 244 & Metabolic network & Undirected & Protein & Interaction \\
Infect-Dublin & 410 & 2765 & Human contact network & Undirected & Person & Proximity \\
Celegans & 453 & 2025 & Metabolic network & Undirected & Substrates & Metabolic reactions \\
Autobahn & 1168 & 2486 & Infrastructure network & Directed & Location & Highway \\
Air traffic control & 1226 & 2615 & Infrastructure network & Directed & Airport/Service center & Preferred route \\
Facebook & 2888 & 2981 & Social network & Undirected & Person & Social relationship \\
Human proteins & 3133 & 6726 & Metabolic network & Undirected & Protein & Interaction \\
Gnutella & 10876 & 39994 & Computer network & Directed & Host & Connection \\
\hline
\end{tabular}}
\end{table*}

\section{Conclusion}

In summary, we proposed a cost-effective network disintegration method called targeted enumeration (TE). Specifically, the TE method was divided into two stages. In the first stage, we used rank aggregation to transform multiple rankings of nodes into a comprehensive ranking. We then selected the top $\tilde{N}$ nodes based on the aggregated ranking as the candidate set of nodes to remove. The size of the candidate set was controlled by the redundancy coefficient $0 \leq \alpha \leq 1$. We showed that rank aggregation can help to find a stable and credible candidate set. The second stage was a targeted enumeration, where, instead of enumerating all possible combinations in the general sense, we enumerated within the scope of the candidate set. The optimal solution was the combination of nodes corresponding to the largest disintegration effect $\Phi$. We showed that a small value of the redundancy coefficient $\alpha$ was sufficient for the targeted enumeration, which is crucial for the feasibility of the TE. Numerical experiments on synthetic and real-world networks have shown that the TE significantly outperforms conventional methods and achieves results that are close to those high-cost intelligent algorithms. In terms of efficiency, the TE was acceptable compared to conventional methods. The critical point of the proposed method was to determine a set of valid candidates. In this study, the introduction of rank aggregation ensured the validity of the candidate set. The aggregated ranking combined multiple node importance criteria and avoided missing potential key nodes from the candidate set. Although it is not the best one in terms of effectiveness or efficiency, the proposed method achieves a satisfying trade-off between effectiveness and efficiency.

The proposed TE method has a highly flexible framework that does not require domain-specific knowledge. Various node importance criteria, rank aggregation methods, and different levels of redundancy coefficient $\alpha$ can be used depending on the real situation. As a typical combinatorial optimization problem, selecting $n$ objects among $N$ objects ($n\ll N$) is common in many application scenarios, including personnel selection, portfolio investment, and drug design. For these problems, finding an optimal solution in a condensed scope is an intuitive approach. The proposed method provides a general executable framework for implementation.

\section*{Acknowledgments}

We thank the teachers and students of the International Academic Center of Complex Systems for their suggestions on this work.


\bibliographystyle{IEEEtran}
\bibliography{wang}

\newpage

\end{document}